\documentclass[aps,twocolumn,showpacs]{revtex4}
\usepackage{graphics}
\usepackage{graphicx}

\begin{document}
\title{Diffusion of photoexcited carriers in graphene}
\author{F.T. Vasko}
\email{fedirvas@buffalo.edu}
\author{V. V. Mitin}
\affiliation{Department of Electrical Engineering, University at Buffalo, Buffalo, NY 14260-1920, USA}
\date{\today}

\begin{abstract}
The diffusion of electron-hole pairs, which are excited in an intrinsic graphene by the ultrashort focused laser pulse in mid-IR or visible spectral region, is described for the cases of peak-like or spread over the passive region distributions of carriers. The spatio-temporal transient optical response on a high-frequency probe beam appears to be strongly dependent on the regime of diffusion and can be used for verification of the elasic relaxation mechanism. Sign flip of the differential transmission coefficient takes place due to interplay of the carrier-induced contribution and weak dynamic conductivity of undoped graphene. 
\end{abstract}

\pacs{72.80.Vp, 73.50.Bk, 78.67.Wj}
\maketitle

The electron energy spectrum of graphene has been characterized from marnetotransport, optical, and photoemission measurements, see reviews \cite{1}, \cite{2}, and \cite{3}, respectively. However, the relaxation and generation-recombination mechanisms are not verified completely at present time. For example, the question of a momentum relaxation mechanism remains controversial in spite of many studies, see review \cite{4} and Refs. 5. An additional information on elastic relaxation can be found from optical response under non-uniform pumping. It was demonstrated recently \cite{6,6a} that a diffusion of photoexcited carriers during short time intervals can be investigated by the all-optical measurements using the focused pump and probe beams which are shifted in the spatio-temporal domain. A theoretical study of the charge diffusion process is not performed yet (only the spin diffusion, which is determined by a slow spin-flip scattering, is investigated \cite{6b}) and the results of Refs. 6 and 7 remain unclear.

In this Letter we study the spatio-temporal evolution of electron-hole pairs after excitation by the focused laser pulse. Such an evolution is governed by the energy-dependent diffusion of carriers caused by the elastic scattering. We analyze the case of weak intercarrier scattering, when a peak-like distribution after mid-IR pumping (the energy relaxation via acoustic phonons is negligible up to nanosecond time scale and the carrier-carrier scattering should not be effective for low and moderate pumping levels) and a distribution spread over the passive region, with energies less than the optical phonon energy, after excitation by a visible pulse takes place. The diffusion processes appear to be different for these cases and it is possible to extract characteristics of elastic scattering in low- and high-energy regions using this approach. The description of the optical response is performed within the local approach if the spatio-temporal scales of carrier evolution exceed the wavelength and period of radiation. It is important, that evolution of the transmission coefficient in the high-frequency region is determined by {\it the spatio-temporal evolution of the carrier energy density} because of the gapless energy spectrum with linear dispersion laws, see Refs. 2b and 9. Due to this, we have analyzed the diffusion processes for distributions of concentration and energy. In addition, the differential transmission (reflection) coefficient depends both on the carrier-induced contribution under investigation and on the weak dynamic conductivity of undoped graphene due to the virtual interband transitions. Such {\it interplay leads to a sign flip} of the relative transmission which depends on probe frequency and on spatio-temporal evolution of photoexcited carriers. Similar phenomenon was analyzed for the case of homogeneous geometry in the response on mid-IR probe. \cite{7} Recently, a sign flip of the spatio-temporal response was observed in the near-IR spectral region \cite{6a} but a mechanism of this peculiarity was not clarified.

The spatio-temporal evolution of carriers after ultrafast non-uniform photoexcitation is desctibed by the same distribution functions for electrons and holes $f_{{\bf xp}t}\simeq f_{{\bf x}pt}+\Delta f_{{\bf xp}t}$. Here we took into account an effective momentum relaxation and separated the weak asymmetric part of distribution, $\Delta f_{{\bf xp}t} \approx -({\bf v}_{\bf p}\cdot\nabla_{\bf x} )f_{{\bf x}pt}/\nu_p$ where $\nu_p$ is the momentum relaxation rate and ${\bf v}_{\bf p}=\upsilon{\bf p}/p$ is the velocity of carriers written through $\upsilon =10^8$ cm/s. The symmetric distribution $f_{{\bf x}pt}$ is governed by the kinetic equation \cite{8} 
\begin{equation}
\frac{\partial f_{{\bf x}pt}}{\partial t}+\overline{({\bf v}_{\bf p}\cdot\nabla_{\bf x})\Delta f_{{\bf xp}t}}=\sum\limits_{r}J_r(f_{{\bf x}t}|p) , 
\end{equation}
where the overline means the averaging over $\bf p$-plane angle and the collision integrals $J_r (f_{{\bf x}t} |{\bf p})$ describe the energy relaxation of carriers via inelastic scattering and the generation-recombination processes. For the ultrafast excitation regime, when the pulse duration $\tau_{ex}$ is shorter in comparison to the diffusion and relaxation times, Eq. (1) should be considered \cite{7} with the initial condition at $t\sim\tau_{ex}\to 0$:
\begin{equation}
f_{{\bf x}p,0}=\nu_{\bf x}\tau_{ex}\Delta\left(\frac{ 2\upsilon p -\hbar\Omega}{\gamma} \right) , ~ \nu _{\bf x}=\frac{\pi}{\hbar\gamma}\left(\frac{e{\cal E}_{\bf x}\upsilon}{\hbar\Omega} \right)^2 .
\end{equation}
Here $\nu_{\bf x}$ is the interband frequency of transitions due to the in-plane electric field ${\cal E}_{\bf x}w_t\exp (-i\Omega t)+$c.c. written through the spatio-temporal modulated field strength ${\cal E}_{\bf x}w_t$ with the temporal form-factor $w_t$ of duration $\tau_{ex}$. The Gaussian form-factor $\Delta (z)=\exp (-z^2)/\sqrt{\pi}$ describes the broadening of interband transitions determined by the phenomenological energy $2\gamma$. 

Performing the averaging over $\bf p$-plane in the left-hand side of Eq. (1), we write the kinetic equation through the $p$-dependent diffusion coefficient $D_p$ as follows
\begin{equation}
\left( {\frac{\partial}{\partial t}-D_p\nabla _{\bf x}^{\bf 2} } \right)f_{{\bf x}pt}=\sum\limits_{r} {J_{r} (f_{{\bf x}t} |p)} , ~~~D_p=\frac{\upsilon^2}{2\nu_p}  .
\end{equation}
Below we fit $D_p$ through the momentum relaxation rate $\nu_p$ determined by the model of scattering by finite- and short-range disorder, $\nu_p =(\upsilon_d p/\hbar ) [\Psi (pl_c/\hbar ) +\upsilon_0/\upsilon_d ]$. \cite{9} Here $\Psi (pl_c/\hbar )$ is a form-factor describing the finite-range scattering with the correlation length $l_c$. The characteristic velocities $\upsilon_d$ or $\upsilon_0$ correspond to finite- or short-range scattering contributions, respectively. Using this fit of $\nu_p$ and the conductivity measurements, \cite{10,11} we plot the diffusion coefficient as it is shown in Fig. 1. Since the density of states and $\nu_p$ vanish at $p\to 0$, one obtains $D_p\propto p^{-1}$ in the low-energy region. In the high-energy region, at $pl_c /\hbar >1$, the dependency of $D_p$ on $\upsilon p$ becomes weaker due to an interplay between finite- and short-range scattering mechanisms. The characteristic energy $\upsilon\hbar /l_c$ is about 83 meV and 125 meV for $l_c=$7.5 nm \cite{10} and 5 nm. \cite{11} The corresponding diffusion coefficients are $D_c\approx$145 cm$^2$/s for parameters of Ref. 10 or $D_c\approx$50 cm$^2$/s and 78 cm$^2$/s for parameters of Refs. 11a and 11b,c respectively. As a result, diffusion coefficient exceeds $10^3$ cm$^2$/s for energies less 50 meV and this estimate should increase for high-mobility samples.
\begin{figure}[ht]
\begin{center}
\includegraphics{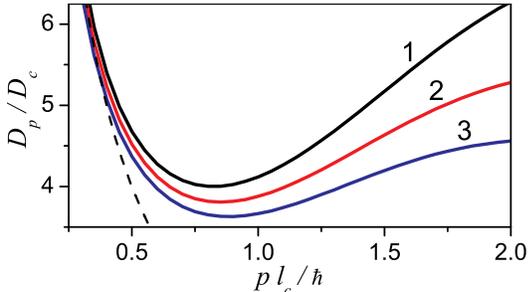}
\end{center}
\addvspace{-0.8 cm}
\caption{(Color online) Dimensionless diffusion coefficient $D_p$ in units 
$D_c =\upsilon^2 l_c /2\upsilon_d$ versus energy $\upsilon p$ in units $\upsilon\hbar /l_c$
for $\upsilon_0 /\upsilon_d =$0.035 (1), 0.05 (2), and 0.065 (3). Dashed curve, $D_p\propto p^{-1}$, corresponds to the short-range scattering case. }
\end{figure} 

For the case of mid-IR pumping, if $\hbar\Omega$ is less than the intra- and intervalley optical phonon energies $\hbar\omega_{\Gamma}$ and $\hbar\omega_K$, the distribution (2) can be used as the initial condition. For the case of pumping in the visible (or near-IR) spectral region, the initial peak (2), formed in the active region $\upsilon p\gg\hbar\omega_K /2$, is transformed into a set of peaks in the passive region after a fast cascade emission of optical phonons. \cite{12} As a result, the form-factor $\Delta (z)$ in Eq. (2) should be replaced by the factor
\begin{equation}
\widetilde{\Delta}\left(\frac{\upsilon p}{\gamma}\right) =
\sum\limits_{k_{\Gamma}k_K}a(k_{\Gamma},k_K)\Delta\left(\frac{\upsilon p-E_{k_{\Gamma}k_K}}{\gamma}\right)\approx\overline{\Delta} .
\end{equation}
Here $k_{\Gamma}$ and $k_K$ are numbers of optical phonons emitted, $E_{k_{\Gamma}k_K}$ is the final energy of peak in the passive region corresponding to the $k_{\Gamma}, k_K$-channel, and the coefficient $a(k_{\Gamma},k_K)$ is determined by the ratio of emission rates for $\Gamma$ and $K$-modes. If $\Omega /\omega_{\Gamma ,K}\geq$3 the number of emission channels exceeds 20 and the initial distribution spreads over the passive region. One can replace $\widetilde{\Delta}(z)$ by the constant $\overline{\Delta}\approx$0.15 - 0.2 if $\gamma\sim$20 meV exceeds the interpeak energies in Eq. (4).

If a diffusion time scale is shorter than the quasielastic energy relaxation scale in the passive region, $\upsilon p<\hbar\omega_K /2$, then we consider the collisionless Eq. (3) with the initial condition (2) or $f_{{\bf x}pt=0}=\nu_{\bf x}\tau_{ex}\overline\Delta$ for the mid-IR or visible pumping cases, respectively. Using the Gaussian initial distribution we write the solution as
\begin{equation}
f_{{\bf x}pt}  = \frac{f_{{\bf x}pt = 0}}{\nu _p^{(D)} t + 1}\exp\left[ -\left( 
\frac{x}{l_{ex}}\right)^2\frac{\nu_p^{(D)}t}{\nu _p^{(D)} t + 1} \right] ,
\end{equation}
where $\nu_p^{(D)}=4D_p/l_{ex}$ is the energy-dependent diffusion rate. For the mid-IR photoexcitation at energy $\hbar\Omega /2$ we use this solution and the definitions for concentration and energy density
\begin{equation}
\left|\begin{array}{*{20}c} n_{{\bf x}t} \\ E_{{\bf x}t}  \\ \end{array}\right| 
=\frac{2}{\pi\hbar^2}\int\limits_0^\infty dpp\left|\begin{array}{*{20}c} 1  \\  
{\upsilon p}\end{array}\right| f_{{\bf x}pt} .
\end{equation}
As a result for the Gaussian distribution of pumping intensity with the lateral size $\sim l_{ex}$, when ${\cal E}_{\bf x}\propto\exp [-(x/l_{ex})^2/2]$ and $\nu_{\bf x}=\nu_R\exp [-(x/l_{ex})^2]$, one obtains the concentration
\begin{equation}
n_{{\bf x}t}=\frac{n_R}{\nu_\Omega t +1}\exp\left[ { - \frac{{\left( {x/l_{ex} } \right)^2 }}{{\nu_\Omega  t + 1}}} \right]
\end{equation} 
and the energy density $E_{{\bf x}t}=\hbar\Omega n_{{\bf x}t}/2$ where we use $\nu_\Omega\equiv \nu_{p=p_\Omega}^{(D)}$ at $p_\Omega =\hbar\Omega /2\upsilon$.
For typical samples, \cite{9,10,11} and CO$_2$ laser pumping with $\hbar\Omega\simeq$140 meV, one obtains the diffusion coefficient $D_{p=p_{\Omega}}=$1300 - 650 cm$^2$/s. The characteristic concentration $n_R$ is about $2\times 10^{12}$ cm$^{-2}$ under pumping intensity $\sim$10 MW/cm$^2$. The spatio-temporal evolutions of $n_{{\bf x}t}$ and $E_{{\bf x}t}$ are identical for the monoenergetic distribution of carriers because the energy relaxation is omitted here.
\begin{figure}[ht]
\begin{center}
\includegraphics[scale=1.1]{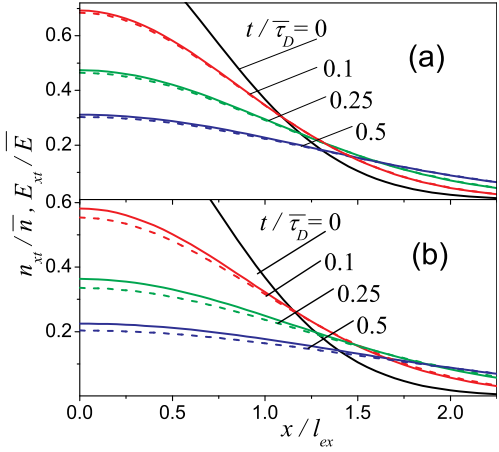}
\end{center}
\addvspace{-0.8 cm}
\caption{(Color online) Normalized distribution of concentration and energy density (solid 
and dashed curves; $\overline{E}\equiv\overline{n}\upsilon\hbar /l_c$) versus dimensionless coordinate $x/l_{ex}$ plotted for the time interval $(0,\tau_D /2)$ and different correlation lengths, $l_c=$5 nm (a) and 10 nm (b). }
\end{figure}

For the case of visible (near-IR) photoexcitation, the numerical integration of (6) should be performed over the passive region, $\upsilon p<\hbar\omega_K /2$. Using the approximation (4), we rewrite the integrals (6) as follows:
\begin{eqnarray}
\left| {\begin{array}{*{20}c}
   {n_{{\bf x}t} }  \\   {E_{{\bf x}t} }  \end{array}} \right|\simeq \overline{n} \left| {\begin{array}{*{20}c} {\Phi _1 (x/l_{ex} ,t/\overline\tau_D )}  \\
{(\upsilon\hbar /l_c )\Phi _2 (x/l_{ex} ,t/\overline\tau_D )} \end{array}} \right| , \nonumber \\
\Phi _k (x,t) = \int\limits_0^{y_m } {\frac{{dyy^k }}{{1 +2t/\chi _y }}} \exp \left( { - \frac{{x^2 }}{{1 +2t/\chi_y }}} \right) , \\
\overline\tau_D =\left(\frac{l_{ex}}{\upsilon}\right)^2\frac{\upsilon_d}{l_c} , ~~~~
\chi_y =\frac{I_1(y^2)}{y^2\exp (y^2)}+\frac{\upsilon_0}{\upsilon_d} . ~ \nonumber
\end{eqnarray}
Here $\overline{n}\simeq 2\nu_R\tau_{ex}\overline{\Delta}/\pi l_c^2$, the relaxation rate $\nu_p$ is written through the first-order Bessel function of an imaginary argument, $I_1(z)$, \cite{9} and the dimensionless cut-off energy is introduced as $y_m=\omega_K l_c/2\upsilon$.

Spatio-temporal evolution of concentration and energy density from the initial Gaussian distribution is plotted in Fig. 2 for typical correlation lengths \cite{10,11} at $\upsilon_0 /\upsilon_d =$0.05. Quenching of energy density from initial Gaussian distribution at $t=0$ becomes faster in comparison to concentration if the correlation length [and $y_m$ in Eq. (8)] increases, c. f. Figs. 2a and 2b. The radial shape of the photoexcited spot transforms from Gaussian to a Lorentz-like distribution and a visible spreading takes place at time scales $\sim 0.1\overline\tau_D$ with $\overline\tau_D$ varying between 3 - 8 ps for the pumping sizes $l_{ex}\simeq$300 - 500 nm. 

The contribution of interband transitions of nonequilibrium carriers (described by the distribution $f_{{\bf x}pt}$) to the response on probe radiation of frequency $\omega$ is described by the addendum to the dynamic conductivity (see Refs. 2b and 9; here $\lambda\to +0$ and $p_\omega\equiv\hbar\omega /2\upsilon$):
\begin{eqnarray}
\Delta\sigma_{{\bf x}t\omega}=\frac{ie^2}{\pi\hbar p_\omega}\int\limits_0^\infty dpp f_{{\bf x}pt}\left(\frac{1}{p_\omega -p+i\lambda} \right. \nonumber \\
\left. -\frac{1}{p_\omega +p+i\lambda}\right) \approx i\frac{e^2\hbar}{\upsilon p_\omega^3} E_{{\bf x}t} .
\end{eqnarray}
The right-hand-side approximation is written for the high-frequency region, where ${\rm Re} \Delta\sigma =0$ and $\Delta\sigma_{{\bf x}t\omega}\propto\omega^{-3}$ is determined through the energy density $E_{{\bf x}t}$ introduced by Eq. (6). It is essential that $\Delta\sigma_{{\bf x}t\omega}$ and transient optical response is determined through energy (not concentration) evolution. Here we used the local approximation because $f_{{\bf x}pt}$ varies slowly over wavelength distances and times $\sim 2\pi /\omega$.
\begin{figure}[ht]
\begin{center}
\includegraphics[scale=1.1]{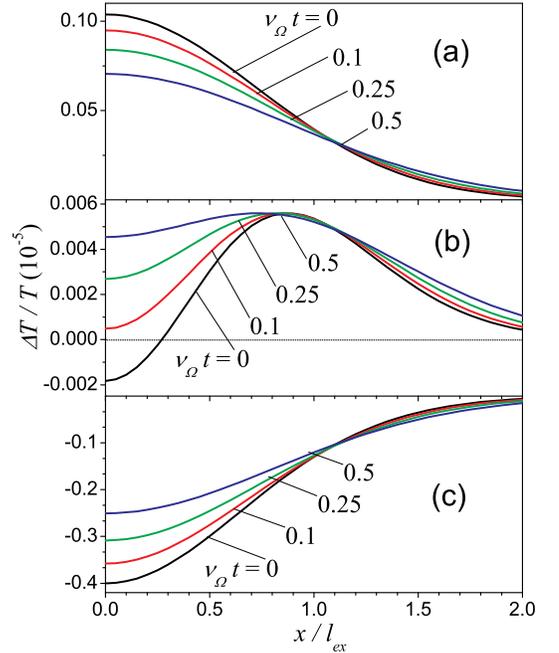}
\end{center}
\addvspace{-0.8 cm}
\caption{(Color online) Spatio-temporal evolution of differential transmission coefficient for the mid-IR pumping case [energy density evolution is determined by Eq. (7)] at probe energies $\hbar\omega =$1 eV (a), 0.78 eV (b), and 0.6 eV (c). }
\end{figure}

Finally, using the results on spatio-temporal evolution of energy density given by Eq. (7) and shown in Fig. 2 we calculate the differential transmission coefficient for the above described regimes of photoexcitation. For the case of normal incidence of probe radiation, the transmission coefficient is given by \cite{2,12,13}
\begin{equation}
T_{{\bf x}t\omega}=\frac{4\sqrt\epsilon}{\left| 1 +\sqrt\epsilon +4\pi\sigma_{{\bf x}t\omega}/c\right|^2 }\simeq T+\Delta T_{{\bf x}t\omega} .
\end{equation}
Below we use $\sigma_{{\bf x}t\omega}\simeq e^2/4\hbar +i\overline{\sigma}_{\omega} +\Delta\sigma_{{\bf x}t\omega}$ written through $\Delta\sigma_{{\bf x}t\omega}$ and a weak contribution to the dynamic conductivity of undoped graphene due to the virtual interband transitions, $\overline{\sigma}_{\omega}$. This contribution is fitted below as $\overline\sigma_\omega\approx (e^2/\hbar )(\varepsilon_m /\hbar\omega -\hbar\omega /\varepsilon_i )$ with the characteristic energies $\varepsilon_m\sim$0.08 eV and $\varepsilon_i\sim$6.75 eV. \cite{13} Within the second-order approach with respect to ${\rm Im}\sigma_{{\bf x}t\omega}$ the differential transmission coefficient takes the form
\begin{equation}
\frac{\Delta T_{{\bf x}t\omega}}{T}\approx -\left(\frac{4\pi}{c}\right)^2
\frac{2\overline\sigma_\omega +{\rm Im}\Delta\sigma _{{\bf x}t\omega}}{(1 +\sqrt\epsilon +\pi e^2 /\hbar c)^2}{\rm Im}\Delta\sigma_{{\bf x}t\omega } .
\end{equation}
Both, $\propto E_{{\bf x}t}^2$ and $\propto E_{{\bf x}t}$, contributions may be important here depending on an interplay between $\overline\sigma_\omega$ and $\Delta\sigma _{{\bf x}t \omega }$ for the high-frequency region described by the right-hand side of Eq. (9). Because of an interplay between $\overline\sigma_\omega$ and $\Delta\sigma_{{\bf x}t\omega}$ contributions to $\Delta T/T$, a sign flip with decreasing of $\omega$ is possible, see below. 
\begin{figure}[ht]
\begin{center}
\includegraphics[scale=1.1]{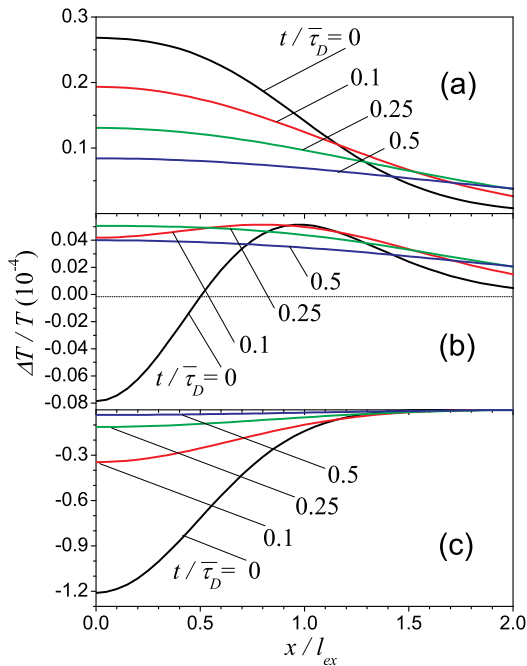}
\end{center}
\addvspace{-0.8 cm}
\caption{(Color online) Spatio-temporal evolution of differential transmission coefficient for the pumping in visible range (energy density evolution is shown in Fig. 2) at probe energies $\hbar\omega =$1.4 eV (a), 1 eV (b), and 0.8 eV (c). }
\end{figure}

Spatio-temporal evolution of the differential transmission coefficient is calculated below for the two regimes of photoexcitation with $E_{{\bf x}t}$ given by Eqs. (7) and (8), respectively. We consider the single-layer graphene with $l_c =$10 nm which is placed on SiO$_2$ with $\sqrt{\epsilon}\simeq 1.46$. In Figs. 3a-c we plot $\Delta T_{{\bf x}t \omega}/T$ for probe energies 1 - 0.6 eV and concentration $n_R\sim 2\times 10^{12}$ cm$^{-2}$. The temporal scale is determined by $\nu_{\Omega}^{-1}\sim$1.6 ps if $l_{ex}\sim$10 $\mu$m. At high frequencies, when $\overline\sigma_\omega <0$, the ratio $\Delta T_{{\bf x}t \omega}/T>0$ as it is shown in Fig. 3a. A weak response with sign flip $\Delta T_{{\bf x}t \omega}/T$ takes place for an intermediate frequency region as it is shown in Fig. 3b and the ratio $\Delta T_{{\bf x}t \omega}/T<0$ for the low frequencies, when $\overline\sigma_\omega >0$. Peak value of $|\Delta T_{{\bf x}t \omega}/T|$ decreases  with frequency before and after the sign flip region and essential spreading of response takes place at $t>0.5/\nu_\Omega$. Since the photoinduced concentration increases under visible excitation (we use here $\overline{n}\sim 10^{14}$ cm$^{-2}$), $|\Delta T_{{\bf x}t \omega}/T|$ also increases up to $10^{-4}$, see Figs. 4a-c plotted for the spectral interval 1.4 - 0.8 eV. With increasing concentration the sign flip of $|\Delta T_{{\bf x}t \omega}/T|$ shifts to higher energies $\hbar\omega\sim$1 eV but the shape of the response is similar to that shown in Fig. 3. Essential spreading of spot takes place at $t>0.3/\overline\tau_D$.

The behavior of $\Delta T_{{\bf x}t \omega}/T$ at different probe frequencies under the near-IR excitation (or the differential reflection coefficient \cite{12}) is in qualitative agreement with the spatio-temporal dependencies of Ref. 7. Numerical estimates for the spatio-temporal scales in Fig. 4 (where $l_{ex}$ and $\overline\tau_D$ are about $\mu$m and ps, respectively; see above) are in good agreement with the experimental data. Thus, no other extra factors should be introduced in order to explain the sign flip peculiarities observed in Ref. 7. But an exact quantitative comparison requires a special treatment of sample parameters (e. g., doping level or parameters of substrate) and additional measurements including the spectral dependencies on pump and probe frequencies. In addition, the approximations used (effect of different relaxation mechanisms on distribution of carriers \cite{7,13} and simplified description of $\overline{\sigma}_{\omega}$ \cite{14} ) should be fitted to experimental conditions. Thus, a special investigation should be performed for a complete study of the diffusion processes under ultrafast optical pumping.

Summarizing, the recent experiment and the theory developed here demonstrate a way for verification of the momentum relaxation rate in the high-energy region, up to the optical phonon energy. These results are also important for the study of heating and saturation effects under an inhomogeneous pumping. \cite{2,13}


\begin{thebibliography}{52}
\bibitem{12}
If the carrier-induced variation of absorption is negligible, the energy conservation gives $\Delta T=-\Delta R$, so that the differential coefficients should be connected by the relation $\Delta R/R=-(\Delta T/T)T/R$.
\end{thebibliography}
\end{document}